\documentclass{appolb}
\usepackage{epsfig}
\begin{document}
\title{Hydrodynamics at RHIC%
\thanks{Presented at Cracow Epiphany Conference on Quarks and Gluons
        in Extreme Conditions, Cracow, Poland, January 2002}%
}
\author{Pasi Huovinen
\address{School of Physics and Astronomy, University of Minnesota,\\
         Minneapolis, MN 55455, USA}
}

\maketitle
\begin{abstract}
The hydrodynamical models used to describe the evolution of heavy-ion
collisions are briefly reviewed and their results compared with
recent RHIC data.
\end{abstract}
  
\section{Introduction}

Hydrodynamical models have certain advantages over transport model
calculations in describing heavy ion collisions. One of the most
important is that, once the equation of state and initial conditions
of the matter are specified, the evolution of the system is
determined. No knowledge of the underlying microscopic processes is
required. This is especially important when studying the predicted
phase transition from hadronic to partonic degrees of freedom (and
vice versa) -- a process for which details are still unknown.

The hydrodynamical description is relatively simple and fulfills the
conservation laws without additional constraints. The use of familiar
concepts like temperature, pressure and flow velocity also provides an
intuitive and transparent picture of the evolution. The price to be
paid for these advantages is a set of bold assumptions: local kinetic
and chemical equilibrium and lack of dissipation. This set of
assumptions may or may not be valid in such a small system as that
formed in a heavy ion collision.

In a hydrodynamical description the evolution is assumed to proceed as
follows: In the initial collision a large fraction of the kinetic
energy of the colliding nuclei is used to create many secondary
particles in a small volume. These particles will collide with each
other sufficiently often to reach a state of local thermal
equilibrium. When the system has reached local equilibrium it is
characterized by the fields of temperature, $T(x)$, chemical
potentials associated with conserved charges, $\mu_i(x)$, and flow
velocity, $u^{\mu}(x)$.  The evolution of these fields is then
determined by the hydrodynamical equations of motion until the system
is so dilute that the assumption of local thermal equilibrium breaks
down and the particles begin to behave as free particles instead.

The numerical solution of hydrodynamical equations of motion in all
three spatial dimensions is a tedious problem. In most approaches some
approximate symmetry is applied to reduce the number of spatial
dimensions where numerical solution is needed to two or one.

In the so-called Bjorken model~\cite{Bjorken} the main idea is the
boost invariance of the longitudinal flow. The longitudinal flow is
assumed to be given by $v_z = z/t$ at all times. This leads to
particularly simple solutions of the equations of motion since the
longitudinal expansion can be solved analytically. Also it is
sufficient to solve the equations of motion in the transverse plane at
$z=0$ since the solution is independent of the boosts along the beam
axis.  The obvious drawback in this approximation is that the
observables are independent of rapidity.

In central collisions of spherical nuclei the expansion can be
simplified by assuming cylindrical symmetry. Of course this symmetry
can not be applied to non-central collisions where the shape of the
source has a crucial role in the buildup of elliptic anisotropy.

\section{Hydrodynamical models}

   \subsection{The basics\protect\footnote{For a more detailed discussion
                                   see ref.~\protect\cite{Rischke:1998}.}}

Hydrodynamics is basically an application of conservation laws. The
local conservation of energy and momentum and any other conserved
four-currents $j_i^\mu$, $i=1,\ldots,n$ are expressed by
\[
   \partial_{\mu}T^{\mu\nu} = 0 {\hspace{1cm}} {\rm and} {\hspace{1cm}}
   \partial_{\mu}j^{\mu}_i = 0,
\]
respectively, where $T^{\mu\nu}$ is energy momentum tensor. Without
any additional constraints these $4+n$ ($n$ is the number of conserved
currents) equations contain $10+4n$ unknown variables. The simplest
and most commonly used approach to close this system of equations is
the ideal fluid approximation which reduces the number of unknown
variables to $5+n$.

In the ideal fluid approximation the energy momentum tensor of the
kinetic theory,
\[
    T^{\mu\nu} = \int \frac{d^3{\bf p}}{(2\pi)^3 E}\,p^\mu p^\nu f(x,{\bf p}),
\]
and currents $j_i$ are supposed to have forms
\[
    T^{\mu\nu} = (\epsilon + p) u^\mu u^\nu - pg^{\mu\nu}
    {\hspace{1cm}} {\rm and} {\hspace{1cm}}
    j^{\mu}_i = n_i u^\mu,
\]
where $\epsilon$, $p$ and $n_i$ are energy density, pressure and
number density of charge $i$ in the local rest frame of the fluid, and
$u^\mu$ is the flow four-velocity of the fluid. In other words all
dissipative effects, such as viscosity and heat conductivity, are
assumed to be zero and the fluid is always in perfect local kinetic
equilibrium. The additional equation needed to close the system of
equations is provided by the equilibrium equation of state (EoS) of
the matter, which connects the pressure to the densities:
$P=P(\epsilon,n_i)$.

In principle it is possible to include small deviations from local
thermal equilibrium by including dissipative effects, but in practice
relativistic viscous hydrodynamics is very difficult to implement and
has not yet been done~\cite{Rischke:1998}. For preliminary results and
estimates of the effects of viscosity, see ref.~\cite{Muronga:2001}.

   \subsection{Equation of state}

The present results from lattice QCD calculations point to a phase
transition from hadronic to partonic degrees of freedom at a
temperature $T_c \approx 155$ -- 175 MeV, but the order of the phase
transition is still uncertain~\cite{Karsch}. So far no nuclear
equation of state (EoS) based on lattice results has been employed in
hydrodynamical calculations, mostly because lattice calculations are
available only at zero net baryon density.

The usual way to construct an EoS is to use the EoS of a hadron
gas at low temperatures and the EoS of an ideal parton gas
with a bag constant at temperatures above the critical temperature
$T_c$ (see \eg ref.~\cite{Sollfrank:1996}). The EoS of an interacting
hadron gas is approximated by the EoS of an ideal resonance
gas with resonances up to 1.5 -- 2 GeV mass. It is known that the
inclusion of higher-lying resonances mimics interactions between
hadrons well in temperatures up to the pion mass~\cite{Prakash}, but
there is no reliable way to check whether this holds at higher
temperatures~\cite{Sollfrank:1996}. The phase boundary is determined
by using the Gibbs criteria. A first order phase transition between
the hadronic and partonic phases is achieved by connecting the EoSs
with the Maxwell construction.

This procedure is thermodynamically consistent and the EoS both
above and below the phase transition temperature is based on well
established models. However, one of its disadvantages is that it is
not possible to determine the phase transition temperature and latent
heat independently. To circumvent this drawback Teaney
\etal\cite{Teaney:2001} took only the speed of sound,
$c_s=\sqrt{\partial p/\partial\epsilon}$, from the bag model EoS and
made the critical temperature and the latent heat explicit parameters
of their model.

Constructing an EoS with a second order phase transition or crossover
between the phases when only the EoSs of the separate phases are known
is nontrivial. To achieve this in a consistent way Zschiesche
\etal\cite{Zschiesche:2001} constructed a family of EoSs based on a
parametrized $\sigma - \omega$ model. By changing the values of the
parameters they were able to create two EoSs with a first order phase
transition but different latent heats and an EoS with a crossover
phase transition. Strictly speaking these EoSs do not contain a
deconfinement phase transition but a chiral phase transition. However
the EoS below and above phase transition temperature is very similar
to the more conventional constructions explained above.

   \subsection{Initialization}
	\label{initials}

Local thermal equilibrium is one of the assumptions of a
hydrodynamical model; the model itself does not specify the mechanism
that leads to an equilibrated state. Since at RHIC energies the
initial particle production is definitely not an adiabatic process,
hydrodynamics can not be used to describe the initial collision. The
hydrodynamical evolution must begin at a sufficient time after the
initial collision when the system has had time to reach thermal
equilibrium. The initial state of the system, \ie the density
distributions and flow velocities at the beginning of the hydrodynamic
evolution, are not given by the model either but must be given as
external input.

When a boost invariant expansion is assumed, the choice of an initial
state is reduced to a choice of transverse density and velocity
profiles. A simple approach is to fix the value of the entropy to
reproduce the observed final particle multiplicity and to distribute
it on the transverse plane assuming a constant density profile within
the radius of the colliding nuclei using a Fermi function to smooth
the edges of the system. However, this approach cannot be applied to
non-central collisions. One must also remember that since it is the
local pressure gradients which drive the development of transverse
flow and the evolution of the system, the details of flow are
sensitive to the details of the initial distributions~\cite{Dumitru}.
In the same way the final anisotropies are proportional to the
deformation of the source and additional constraints to the initial
distributions are required.

It is known that up to SPS energies the multiplicity
scales with the number of nucleons participating in the
collision~\cite{WA98}. On the other hand, in the high energy limit one
expects the individual parton-parton collisions to contribute equally
to primary particle production and therefore the multiplicity should scale
with the number of binary collisions~\cite{K-tie}. Therefore it is
natural to initialize the system using a localized version of these
approaches: to assume that the density is proportional either to the
number of participants or to the number of binary collisions per unit
area in the transverse plane. Both approaches, or a combination of them
can be used to fix the initial entropy or energy density; a comparison
was made in ref.~\cite{Kolb:2001}. The initial transverse flow
velocity is customarily assumed to be zero, although pre-equilibrium
density gradients might lead to small, but finite, transverse flow
velocities at the time of thermalization.

If the assumption of boost invariance is relaxed the choice of
initial state becomes considerably more complicated. There are
few constraints for the flow velocity profile or the longitudinal
density distributions. The choice of a particular parametrization
and the values of the parameters is largely based on trial and
error -- tuning the model until a reasonable fit to experimental
rapidity distributions is achieved. Even for the same EoS there
are several possible initial states which lead to an acceptable
reproduction of the data~\cite{Huovinen:1998}. For a sample of
initial profiles used successfully see
refs.~\cite{Sollfrank:1996,Ornik:1996,Sollfrank:1998,Hirano}.

An alternative approach to determine the initial state is to use some
other model to calculate it. For example, event generators~\cite{Schlei:1998}
or perturbative QCD (pQCD) calculations~\cite{Eskola:2001} have been
used for this purpose. Even if these approaches increase the predictive
power of hydrodynamics, thermalization is still an additional assumption.

   \subsection{Freeze-out}

At some point in the evolution particles will begin to behave as free
particles instead of a fluid and the hydrodynamical description breaks
down. When and where that happens is not given by hydrodynamics but
must be included as an external input. The conventional approach is to
assume this to take place as a sudden transition from local thermal
equilibrium to free streaming when the mean free path of the particles
becomes larger than the system size, or when the expansion rate of the
system is larger than the collision rate between particles. Finding
where these conditions are fulfilled is a nontrivial problem. Since
the mean free path is strongly dependent on temperature the usual
approximation assumes that the freeze-out takes place on a
hypersurface where temperature (or energy density) has a chosen
freeze-out value. This temperature is of the order of the pion mass,
but its exact value is largely a free parameter which can be chosen to
fit the data. In Pb+Pb collisions at the SPS the values of freeze-out
temperatures vary between 100 and 140 MeV in different
calculations~\cite{Rischke:2001}.

These values are somewhat smaller than the $T\sim 160$ MeV freeze-out
temperatures obtained using thermal models to fit the particle
abundancies at final state~\cite{Cleymans}. This can be understood by
noticing that the former approaches assume kinetic equilibrium while
the thermal models assume chemical equilibrium. Since chemical
equilibrium requires frequent inelastic collisions, while kinetic
equilibrium only requires elastic collisions, it is natural to assume
that inelastic collisions cease first and chemical freeze-out occurs
at higher temperatures than kinetic freeze-out. Thus the system may be
in local kinetic, but not chemical, equilibrium at the later stages of
its evolution. How much this would affect the EoS and whether this
change in the EoS would have any observable effects in the evolution
of the system is so far largely unexplored (with some early exceptions
like ref.~\cite{Bebie}).  However, first preliminary results have been
shown and more are in preparation~\cite{Hirano:2002}.

After choosing the surface where the freeze-out takes place, the
thermodynamic variables characterizing the state of the fluid must be
converted to spectra of observable particles. A practical way of doing
this is the Cooper-Frye algorithm~\cite{Cooper-Frye} where the
invariant momentum distribution of a hadron $h$ is given by
\[
  E \frac{dN}{d^{3}p} 
    =  \frac{g_h}{(2 \pi)^{3}} \int_{\sigma_{f}} 
          \frac{1}{exp[(p_{\mu}u^{\mu}-\mu)/T] \pm 1} \,
                           p^{\mu} d\sigma_{\mu}.
\]
Here the temperature $T(x)$, chemical potential $\mu(x)$ and flow
velocity $u^{\mu}(x)$ are the values on the decoupling surface
$\sigma_{f}$. Besides its relative simplicity, this approach has the
advantage that if the same equation of state is used on both sides of
decoupling surface, both energy and momentum are conserved.  However,
the Cooper-Frye formula has a conceptual problem. At those areas where
the freeze-out surface is spacelike, the product $p^{\mu}
d\sigma_{\mu}$ may be either positive or negative, depending on the
value and direction of $p^{\mu}$. In other words, the number of
particles freezing out on some parts of the freeze-out surface may be
negative. These negative contributions are small (few per cent, see
ref.~\cite{Teaney:2001}) and usually ignored. More refined procedures
without negative contributions have been suggested~\cite{Bugaev} but
their implementation is complicated. A model using one of these
refined freeze-out procedures is in preparation~\cite{Csernai} and it
will be interesting to see how large an effect the freeze-out
procedure has on the final particle spectra in a full-fledged
calculation.

Another way to refine the hydrodynamical freeze-out procedure is to
switch from a hydrodynamical to a microscopic transport model
description well within the region where hydrodynamics is supposed to
be applicable~\cite{Teaney:2001,Bass}.  Besides giving a better
description of freeze-out, such models include the separate chemical
and kinetic freeze-outs. The main drawback of such models -- apart
from the increased complexity -- is that the region where the switch
from hydro to transport description should take place is as poorly
defined as kinetic freeze-out surface in ordinary hydrodynamical
calculation. The educated guess employed in both
refs.~\cite{Teaney:2001,Bass} is that the switch happens immediately
after hadronization.

\section{Comparison with the data}

  \subsection{Charged particle multiplicity and transverse energy}

\begin{figure}
 \begin{minipage}{61mm}
  \epsfxsize 61mm \epsfbox{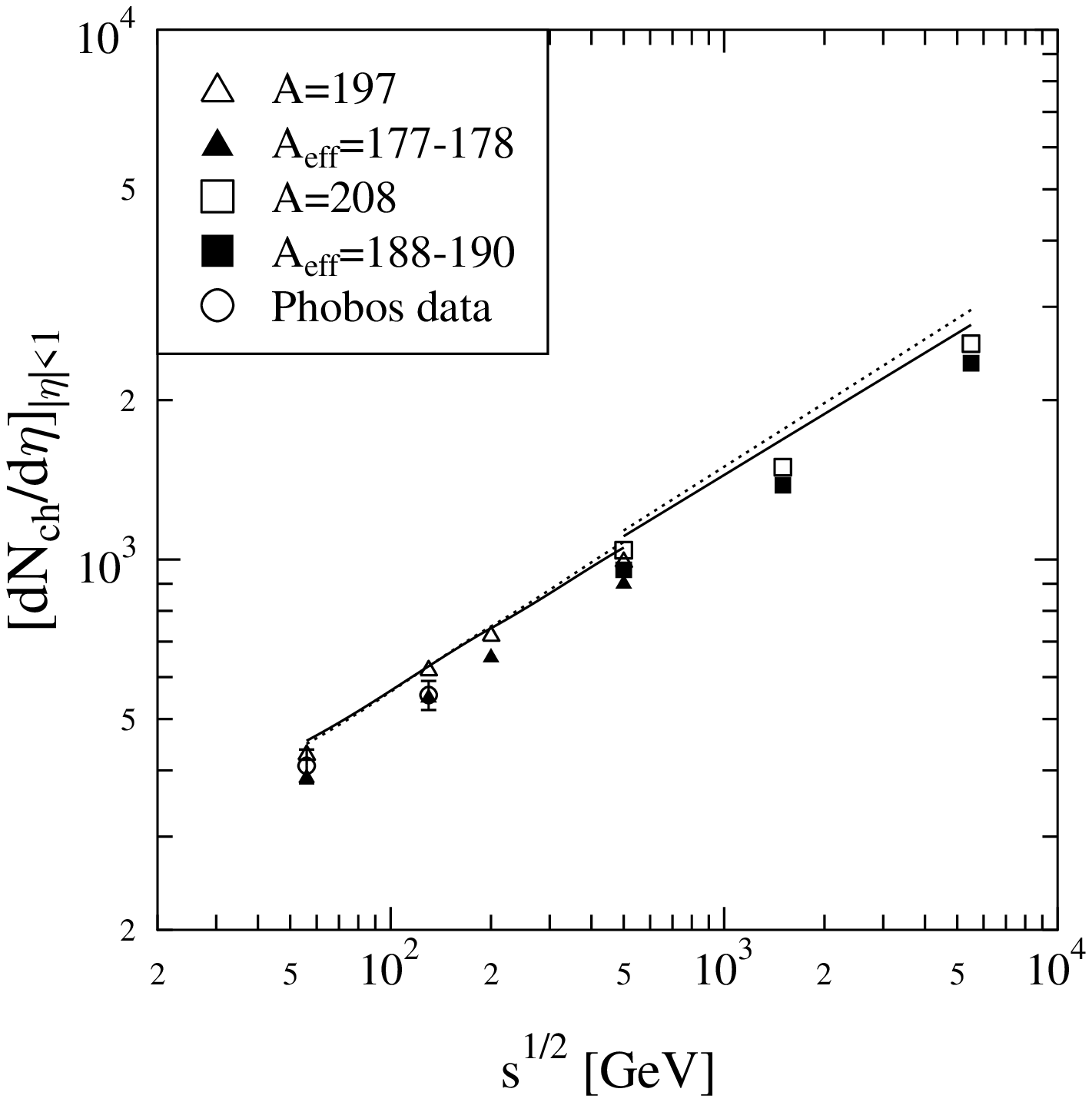}
  \caption{Charged particle multiplicity
           calculated at different collision
           energies and compared to Phobos data~\cite{Back:2000}
           and calculations (solid and dotted lines)
           in ref.~\cite{EKRT}. A=197 and 208 refer to the mass
           of the nuclei and A$_\mathrm{eff}$ to a finite impact parameter.
	   Figure and calculation are from ref.~\cite{Eskola:2001}.}
  \label{saturation}
  \end{minipage}
	\hfill
  \begin{minipage}{61mm}
    \epsfxsize 61mm \epsfbox{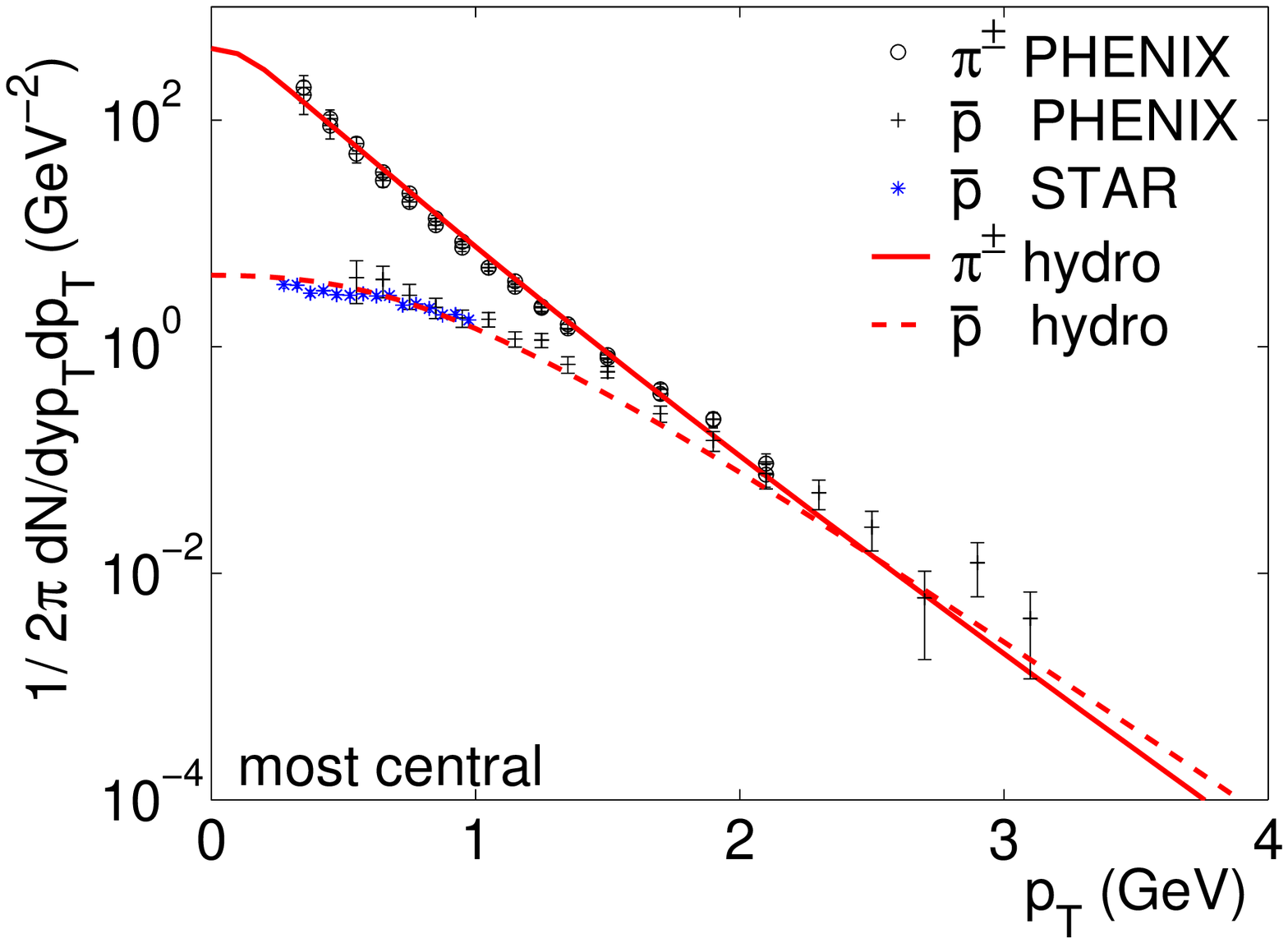} 
    \caption{Charged pion and antiproton spectra from most central Au+Au
             collisions at the $\sqrt{s_{NN}} = 130$ GeV as measured by the
             Phenix~\cite{Velkovska:2001} and STAR~\cite{Adler:2001}
             collaborations. The hydrodynamical calculation and figure are
             from ref.~\cite{Heinz:2001}.}
    \label{pbarpi}
  \end{minipage}
\end{figure}

In ideal fluid hydrodynamics the final particle multiplicity
is proportional to the entropy of the initial state. Thus in a
certain sense multiplicity is not really a \emph{hydrodynamical}
result, but something one dials in as input. Nevertheless, when the
initial state is not chosen by comparing the final result to data, but
calculated from perturbative QCD as in ref.~\cite{Eskola:2001},
the final particle multiplicity becomes a prediction of the model.

In ref.~\cite{Eskola:2001} Eskola \etal used the EKRT saturation model
to calculate the production of minijets in the primary collisions and
then converted this result to the initial state of hydrodynamic
evolution. Their results for charged particle multiplicities at RHIC
and LHC energies are shown in fig.~\ref{saturation}. The results agree
well with the data measured by the Phobos
collaboration~\cite{Back:2000,Back:2001}. It is important to remember
that no fitting or fine tuning was done to achieve this result, and
that the multiplicity at $\sqrt{s_{NN}}=200$ GeV energy was not yet
measured when this calculation was done.

When compared with the EKRT saturation results shown in ref.~\cite{EKRT},
which did not contain an expansion stage, one notices that hydrodynamical
expansion causes only a small change in multiplicity, but the transverse
energy decreases by a factor 3. However, at RHIC the calculated $E_t$
is still 10--20 \% larger than the experimental value~\cite{Phenix-Et}.

  \subsection{$p_T$ spectra}

The conventional way to initialize a hydrodynamical calculation is to
use experimental hadron data to fix the initial values. Depending on
the details of the initialization the calculated $p_t$ distributions
can be only fits to the data or have some predictive power. In
ref.~\cite{Heinz:2001} the initialization is done by first choosing
the initial entropy and the net baryon densities to reproduce the
observed pion multiplicity and $\bar{p}/p$ ratio in the most central
collisions. Then a combination of different parametrizations mentioned
in section~\ref{initials} and ref.~\cite{Kolb:2001} is chosen to
reproduce the observed multiplicity per participant as function of
centrality. In this process the slopes of the spectra or the
centrality dependence of the slopes are not used as an input and their
calculated values can be taken as predictions.

In central collisions this initialization led to a maximum
initial temperature $T_\mathrm{max}=328$ MeV and an energy density
$\epsilon_\mathrm{max}=21.4$ GeV/fm$^3$ at an initial time $\tau_0=0.6$
fm/$c$.  At time $\tau=1$ fm/$c$, usually used to estimate initial
energy density via Bjorken's formula~\cite{Bjorken}, the corresponding
average energy density is $\langle\epsilon\rangle = 5.4$ GeV/fm$^3$, which
is consistent with experimental estimates~\cite{Phenix-Et}.

\begin{figure}
  \begin{minipage}{61mm}
    \epsfxsize 61mm \epsfbox{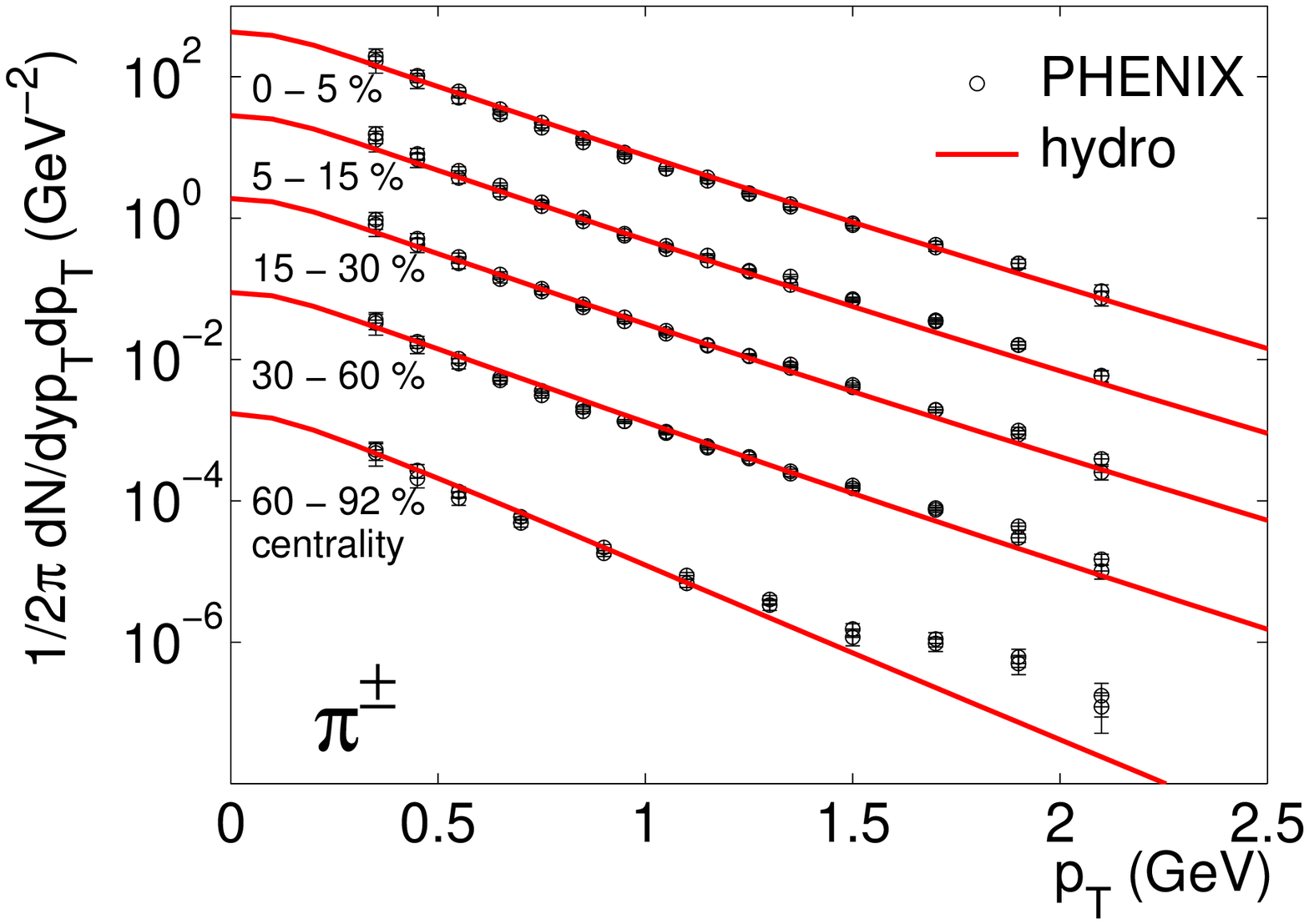} 
  \end{minipage}
	\hfill
  \begin{minipage}{61mm}
    \epsfxsize 61mm \epsfbox{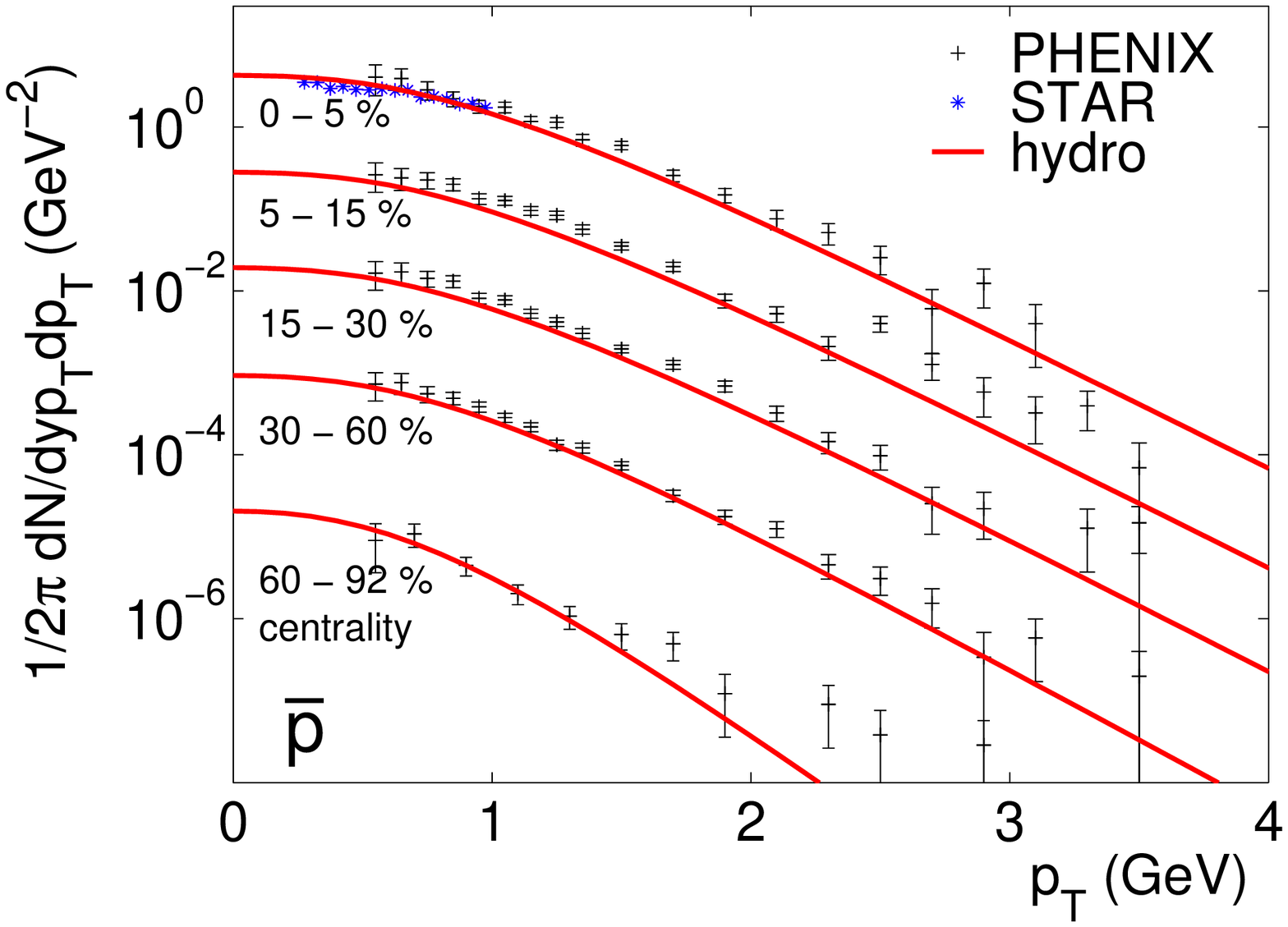} 
  \end{minipage}
    \caption{Charged pion (left panel) and antiproton (right panel) 
             spectra from Au+Au collisions at $\sqrt{s_{NN}} = 130$ GeV at
             various centralities as measured by the
             Phenix~\cite{Velkovska:2001} and STAR~\cite{Adler:2001}
             collaborations. The hydrodynamical calculation and figures are
             from ref.~\cite{Heinz:2001}.}
    \label{centrality}
\end{figure}

Local chemical equilibrium is assumed to hold until kinetic
freeze-out. Thus chemical and kinetic freeze-outs take place at the
same temperature. Even if the pion yield is correctly reproduced
in ref.~\cite{Heinz:2001}, it is not possible to obtain both correct
proton and antiproton yields simultaneously. The authors have chosen
to circumvent this problem by calculating the particle yields at
hadronization temperature $T_c=T_\mathrm{chem}=165$ MeV and
calculating the slopes of the $p_t$ spectra at $T_f=128$
MeV. Subsequently they rescale the particle yields at kinetic
freeze-out to their values at chemical freeze-out by hand. The results
shown here in figs.~\ref{pbarpi} and~\ref{centrality} are very similar
to those in ref.~\cite{Teaney:2001}.  Since separate chemical and
kinetic freeze-out was included in the transport model description of
hadronic phase of ref.~\cite{Teaney:2001}, one can conclude that the
method applied in ref.~\cite{Heinz:2001} provides a reasonable
approximation.

It has been observed that around $p_t=2$ GeV the antiproton and pion
yields are roughly equal, and the slopes of the distributions suggest that
the antiproton yield is \emph{larger} than pion yield for $p_t > 2$
GeV~\cite{Velkovska:2001}. As can be seen in fig.~\ref{pbarpi}
this phenomenon can be explained as
a simple consequence of strong transverse flow.
Besides giving an explanation for this so-called anomalous $\bar{p}/\pi$
ratio, hydrodynamical calculation provides very good fit to both pion
and antiproton spectra both in central and semicentral collisions. As
can be seen in fig.~\ref{centrality} the deviation from the data
is significant only in the most peripheral collisions.

  \subsection{Elliptic anisotropy}
	\label{v2}

Since the initial particle production is azimuthally symmetric,
azimuthal anisotropy of the final particle distributions is a signal of
rescatterings among produced particles. More frequent rescattering can
be expected to lead to a larger anisotropy and since hydrodynamics
assumes zero mean free path and thus an infinite scattering rate, it
provides an upper limit to observable anisotropies. Anisotropy is
quantified by measuring the harmonic coefficients $v_n(y,p_t;b)$ of a
Fourier expansion in $\phi_p$ of the measured hadron spectrum
$dN/(dy\,p_t\,dp_t\,d\phi_p)$~\cite{Voloshin}. Anisotropy
characterized by a non-zero second coefficient, $v_2$, is called
elliptic anisotropy or elliptic flow~\cite{Ollitrault}.

\begin{figure}
  \begin{center}
    \epsfxsize 65mm \epsfbox{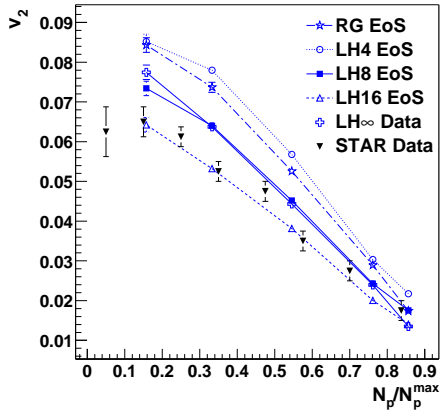}
  \end{center}
    \caption{Elliptic anisotropy coefficient $v_2$ for charged particles
             at RHIC as a function of the number of participants (relative
             to the maximum) as measured by the STAR
             collaboration~\cite{Ackermann:2000} and
             calculated using different equations of state~\cite{Teaney:2001}.
             The label RG corresponds to a resonance gas EoS without a phase
             transition. The number in other labels describes the latent heat.
             The figure is from ref.~\cite{Teaney:2001}.}
    \label{v2b}
\end{figure}

In fig.~\ref{v2b} the centrality dependence of the elliptic anisotropy
coefficient $v_2$ in Au+Au collisions at $\sqrt{s_{NN}}=130$ GeV
energy is shown~\cite{Ackermann:2000}. The calculations of
refs.~\cite{Teaney:2001,Kolb:2000} give very similar results: in
central and semicentral collisions the data reaches the hydrodynamical
limit and depending on the EoS and freeze-out temperature (the latter
is not shown in fig.~\ref{v2b}) the agreement is satisfactory even
close to peripheral collisions. The calculation shown in
fig.~\ref{v2b} was carried out using various EoSs with different
latent heats~\cite{Teaney:2001}. The best fit to the data is obtained
using an EoS with a relatively large latent heat, but when other
observables are considered, the authors conclude that LH8 EoS with
smaller latent heat leads to best overall description of data.

In a hydrodynamical model the final anisotropy is proportional to
the initial deformation of the source. The deformation as a function of
impact parameter depends on the particular parametrization applied
but, as shown in ref.~\cite{Kolb:2001}, when the anisotropy is averaged
over all momenta and centrality is expressed as a fraction of the
total multiplicity, differences between parametrizations even out. Thus
the result shown in fig.~\ref{v2b} is independent of the particular
parametrization.

\begin{figure}
  \begin{center}
    \epsfxsize 70mm \epsfbox{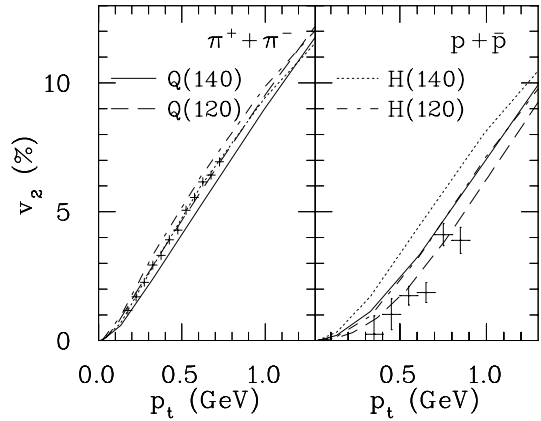}
  \end{center}
    \caption{$p_t$ differential elliptic anisotropy of pions (left panel)
             and protons+antiprotons (right panel) in minimum bias
             collisions as measured by the STAR collaboration~\cite{Star:v2pt}
             and calculated using different equations of state and
             freeze-out temperatures~\cite{Huovinen:2001}. The letters
             Q and H in the labels stand for an EoS with a first order
             phase transition and a hadron gas EoS without a phase
             transition, respectively. Numbers in parentheses stand
             for the freeze-out temperature in MeV.}
    \label{v2pt}
\end{figure}

The $p_t$ differential anisotropy, $v_2(p_t)$, for pions and for the
sum of protons and antiprotons~\cite{Star:v2pt} depicted in
fig.~\ref{v2pt} also shows clear hydrodynamical behaviour. As
predicted\footnote{A similar result was later obtained in
                   ref.~\cite{Teaney:2001}.}
in ref.~\cite{Huovinen:2001}, the heavier the particle, the smaller the
anisotropy at low $p_t$. So far there are no published data regarding the
$p_t$ differential anisotropy of strange particles so whether they
also obey this rule remains to be seen. The identified particle data
is so far limited to the low $p_t$ region shown in fig.~\ref{v2pt}.
The $p_t$ differential anisotropy of negative hadrons~\cite{Snellings}
follows the hydrodynamical calculations up to transverse momenta $p_t
\le 1.5$ -- 2 GeV (not shown) where the anisotropy saturates. The
deviation can be understood as a sign of incomplete thermalization of
high-$p_t$ particles.

It is well known that the changes in the EoS can be compensated by
changing the initial state and freeze-out temperature (see \eg
ref.~\cite{Huovinen:1998}). A remarkable feature of fig.~\ref{v2pt} is
that changes in the EoS affect the anisotropy of both pions and
nucleons in the same way -- a stiffer EoS leads to a larger anisotropy
at low values of $p_t$ -- but a change in freeze-out temperature
changes the pion and nucleon anisotropies in opposite
directions. A lower freeze-out temperature leads to a larger
anisotropy for pions but to a smaller anisotropy for nucleons (for
further discussion, see ref.~\cite{Huovinen:2001}). This may provide
an additional method of constraining possible equations of state and
freeze-out temperatures.

\begin{figure}
 \begin{center}
  \epsfxsize 80mm \epsfbox{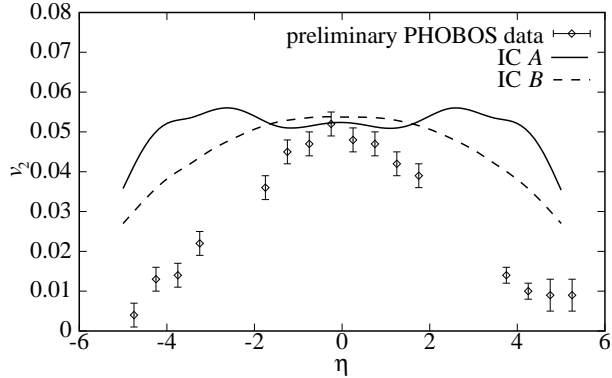}
 \end{center}
  \caption{Pseudorapidity dependence of elliptic anisotropy for charged
           particles in minimum bias collisions at $\sqrt{s_{NN}}=130$
           GeV energy. The figure and calculations are from ref.~\cite{Hirano}
           and the experimental data from ref.~\cite{Phobos-v2}. Dashed
           and dotted lines correspond to different initializations of
           the model, see ref.~\cite{Hirano}.}
  \label{v2eta}
\end{figure}

In the studies mentioned above the expansion was assumed to be boost
invariant. Judging by the measured rapidity
distributions~\cite{Phobos-dn}, this is a reasonable assumption close
to midrapidity, but it makes it impossible to make any statements
about the rapidity dependence of any variable. To study the rapidity
dependence of the elliptic anisotropy the assumption of boost
invariance was relaxed and the calculation was done using a genuinely
three dimensional model in ref.~\cite{Hirano}. When compared to the
excellent agreement with the data in figs.~\ref{v2b} and~\ref{v2pt},
the result depicted in fig.~\ref{v2eta}~\cite{Hirano} may look less
satisfactory. The data~\cite{Phobos-v2} reaches the hydrodynamical
value only around midrapidity. On the other hand, even this
result reproduces the data within a one to two units of pseudorapidity
wide window. This area already contains most of the produced
particles. It is also worth remembering that anisotropy in
hydrodynamical models depends strongly on the initial shape of the
system. The initialization used in ref.~\cite{Hirano} is relatively
simple and more sophisticated initialization may lead to better fit to
the data. Therefore it is premature to conclude based on this data and
calculation alone that thermalization is reached only at midrapidity

  \subsection{Two particle correlations}

Two particle momenta correlations, known as HBT interferometry,
provide a method to study the space-time structure of the emitting
source~\cite{HBTreview}. It has been predicted that a
first order phase transition would lead to unusually large
HBT-radii~\cite{Rischke:1996}.  However, comparisons of
calculations~\cite{Zschiesche:2001,Heinz:2001,Soff} with
data~\cite{STAR-HBT,Phenix-HBT} have lead to the so-called
\emph{HBT-puzzle}: All calculations give a ratio of HBT-radii
$R_\mathrm{out}/R_\mathrm{side}$ larger than one, but the
experimental value is of order one. The calculated values of
$R_\mathrm{out}$ are also larger and, with the exception of
ref.~\cite{Soff}, values of $R_\mathrm{side}$ are smaller than
observed.

\begin{figure}
 \begin{minipage}{62mm}
  \epsfxsize 62mm \epsfbox{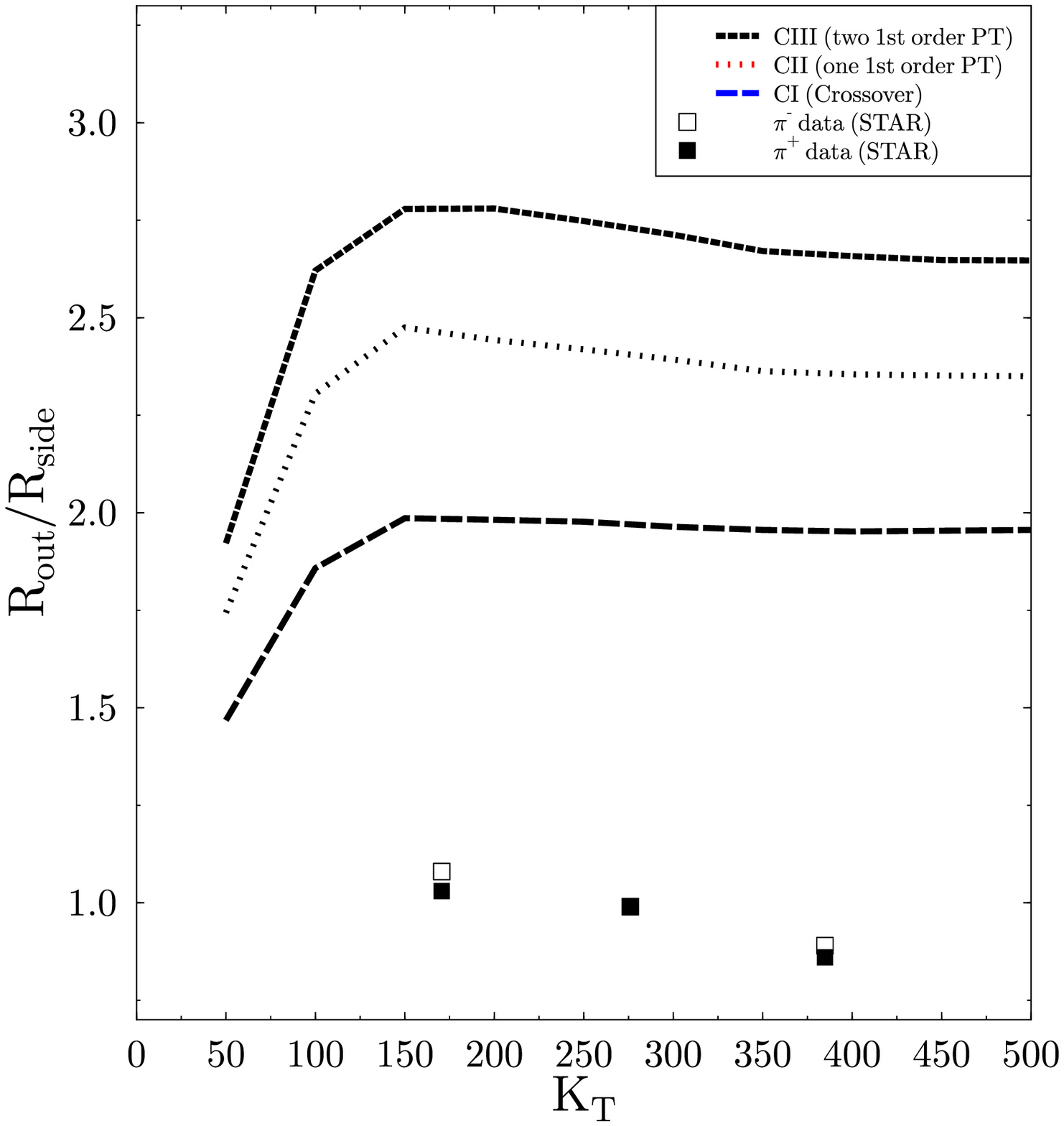}
 \end{minipage}
	\hfill
 \begin{minipage}{62mm}
  \epsfxsize 62mm \epsfbox{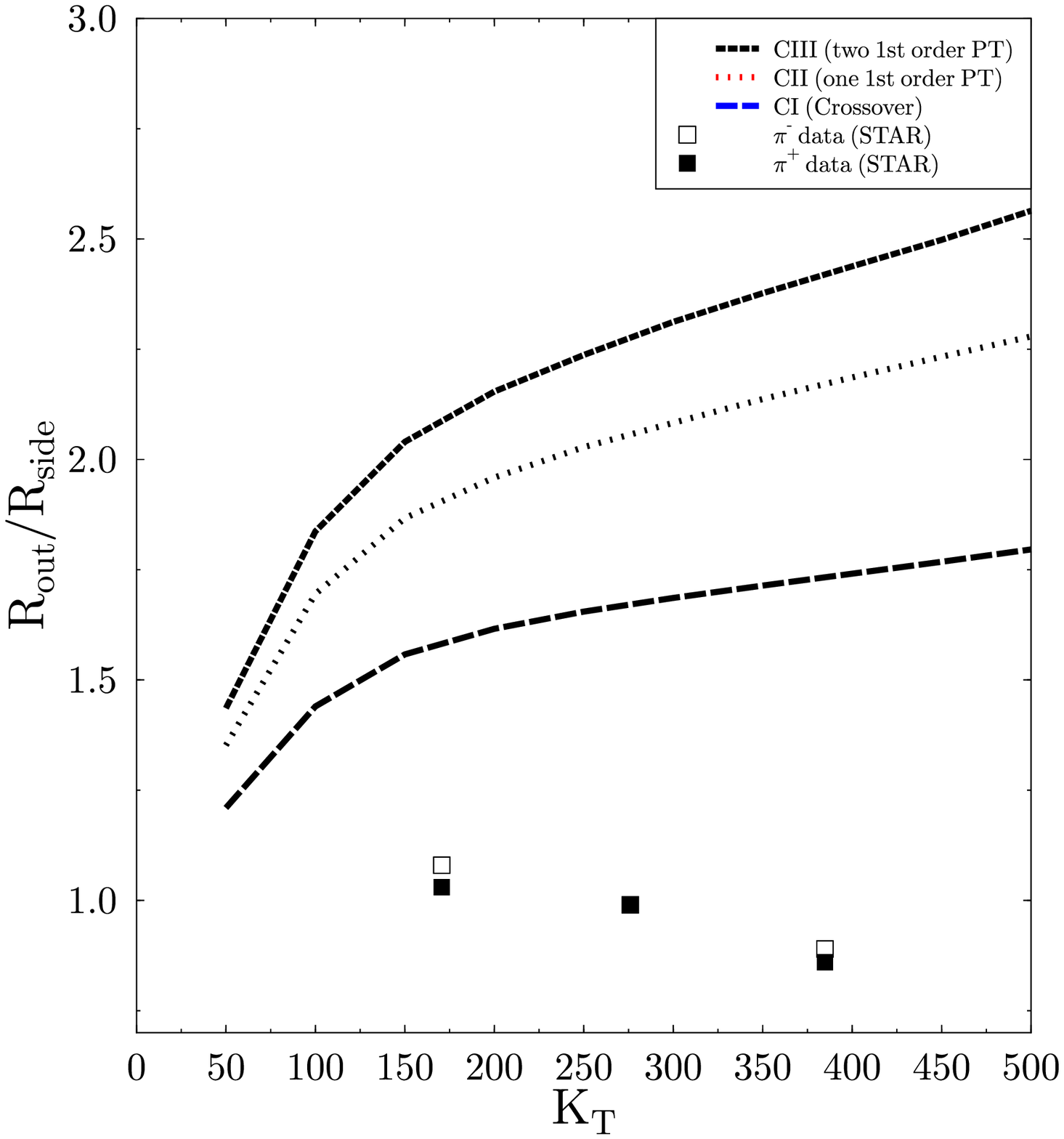}
 \end{minipage}
  \caption{R$_\mathrm{out}$/R$_\mathrm{side}$ as function of $k_t$
           for $T_f=130$ (left panel) and 80 MeV (right panel) using
           different equations of state compared to STAR
           data~\cite{STAR-HBT}. The calculation and figures
           are from ref.~\cite{Zschiesche:2001}.}
  \label{rout-rside}
\end{figure}

It has been suggested that the solution to this puzzle lies primarily
in the description of freeze-out~\cite{Heinz:2001}. This is doubtful
since even calculations where the hadronic phase is described using a
transport model cannot describe the data correctly~\cite{Soff}, even
though this kind of freeze-out description should be more
reliable. Some other theoretical uncertainties, such as the order of
the phase transition and choice of freeze-out temperature, were
addressed in ref.~\cite{Zschiesche:2001}. As shown in
fig.~\ref{rout-rside}, it was found that the freeze-out temperature
has quite a small effect on the radii nor were any of the tested EoSs
(EoS with strong first order, weak first order and cross-over phase
transitions) able to provide an acceptable reproduction of the
data. Since the EoS with a smooth cross-over phase transition (CI in
fig.~\ref{rout-rside}) is closest to the data, it is possible to claim
that HBT measurements favour an EoS with smooth cross-over. This is
particularly interesting since, as argued in section~\ref{v2},
elliptic anisotropy seems to favour a moderately strong first order
phase transition. The explanation to this seemingly contradictory
behaviour, as well as to the entire HBT-puzzle, is still unknown at
present.

\section{Summary}

Hydrodynamical models have been very successful in explaining the
single particle RHIC data at low $p_t$. The $p_t$ spectra and
anisotropies in central and semicentral collisions are well reproduced
for $p_t \le 1.5$ -- 2 GeV and the $\bar{p}/\pi$ ratio at $p_t\sim 2$
GeV/$c$ has a simple explanation due to flow. Especially impressive
has been how hydrodynamics is able to create simultaneously elliptic
anisotropy of negative hadrons which is large enough and anisotropy of
protons which is small enough to fit the data. If one considers
solely this data the collision system behaves
like a thermal system.

However, the reproduction of the HBT-radii has been unsuccessful so
far. It is unclear whether one should refine the final freeze-out
process, hadronization process, or initial state to reach an acceptable
description of the data. Especially puzzling is the fact that the
HBT-radii seem to favour a relatively stiff equation of state with a
crossover phase transition, whereas elliptic anisotropy of protons
favours a soft equation of state with a first order phase transition.

\section*{Acknowledgements}

\noindent
I thank the organizers for the invitation to this conference and
P.~J.~Ellis, J.~I.~Kapusta and A.~Muronga for careful readings of
the manuscript. This work was supported by the US Department of
Energy grant DE-FG02-87ER40328.

\end{document}